\documentclass{JHEP3}

 \usepackage{graphics}
 \usepackage{graphicx}
 \usepackage{epsfig}

\usepackage{amssymb}
\usepackage{amsmath}

\newcommand{\bdm}  {\mathcal D}
\def \to {\rightarrow}

\title{An Iterative Rejection Sampling Method} 

\author{A.~Sherstnev
\\
Cavendish Laboratory, University of Cambridge, \\JJ Thomson Avenue, Cambridge, CB3 0HE, UK
\\
and
\\
Scobeltsyn Institute of Nuclear Physics of Lomonosov Moscow State 
University, \\Moscow, Russia, 119992 (on leave)
}

\abstract{
In the note we consider an iterative generalisation of the rejection sampling 
method. In high energy physics, this sampling is frequently used for event 
generation, i.e. preparation of phase space points distributed according to 
a matrix element squared $|M|^2$ for a scattering process. In many realistic 
cases $|M|^2$ is a complicated multi-dimensional function, so, the standard 
von Neumann procedure has quite low efficiency, even if an error reducing 
technique, like VEGAS, is applied. As a result of that, many of the 
$|M|^2$ calculations go to ``waste''. The considered iterative modification 
of the procedure can extract more ``unweighted'' events, i.e. distributed 
according to $|M|^2$. In several simple examples we show practical benefits 
of the technique and obtain more events than the standard von Neumann method, 
without any extra calculations of $|M|^2$. 
}


\preprint{Cavendish-HEP-08/10}

\begin{document}



\section{Introduction}
\label{intro}
Simulated events of particle scattering are one of the main tools both in 
modern theoretical research and in current experimental analyses in the 
high energy physics (HEP). Usually the events are prepared by means of 
Monte Carlo techniques. The reason for the success of the methods is very 
clear: current simulation problems require realistic events of the processes 
$2\to N$, where N is set by the accelerator energy and is equal to several 
hundreds for the LHC, for instance. But even applied approximations such 
as partons showers and zero width decays for the final particles leave us 
with $N\approx 3-10$. This large number corresponds to a $5-25$ dimension 
phase space. Owing to the large number of dimensions and complexity of the 
investigated functions -- the matrix elements squared -- we cannot rely on 
deterministic grid methods due to the well-known ``curse of dimensionality''. 

The standard scheme applied is the following: at first we prepare a matrix 
element squared symbolically (by hand or automatically), construct an 
importance sampling function, again manually or automatically by applying, 
for example, VEGAS~\cite{Lepage:1977sw}; obtain an estimation of the total 
integral (the total cross section), for chosen cuts and parameters, by the 
classic Monte-Carlo method, find the function maximum (or an array of maxima, 
since stratified sampling may be worth applying), generate enough weighted 
events for the function, and perform the ``hit-and-miss'' von Neumann 
procedure~\cite{vonNeumann:1951}. The final step maps the weighted points, 
distributed according to the importance sampling function, to a subset of 
the sample -- realistic independent unweighted events distributed according 
to the matrix element squared. This is the general and oversimplified scheme 
of functioning of almost all modern Monte-Carlo generators, see for example 
the review~\cite{Dobbs:2004qw}. Useful information on  current mathematical 
Monte-Carlo techniques applied in HEP can also be found in the short review 
by Weinzierl~\cite{Weinzierl:2000wd}. 

Although the scheme described guarantees that the unweighted events are 
distributed according to the function considered, the selection efficiency 
usually is not too high, especially if the function has lots of sharp peaks. 
The main reason is that importance sampling functions are simple functions 
and they have much smoother behaviour and describe the real function 
behaviour in a very crude approximation. As a result we get lots of 
calculated points going into the ``waste bin''. Since the cost calculating 
of the investigate function can be quite large it would be worth considering 
whether we can extract more information (e.g. more unweighted events) from 
the ``waste''. The most obvious way to achieve this goal is to repeat the 
von Neumann procedure with the rejected phase space points. It is rather 
clear that these points {\it can} be used for the task, but with worse 
quality. Since, at the first rejection stage, some points have been accepted 
and, so, removed from the sample of the weighted events, we should change 
the weights of all remaining points. The paper is devoted to the construction 
of a procedure, which gives us the possibility to use the rejected points 
iteratively. 

In fact, the idea of using rejected points is not quite new. For example, 
an iterative von Neumann procedure for the extraction of random bits from 
an independent but biased bit array was constructed in~\cite{Peres:1992}. 
It originates from one of von Neumann's ideas published in his famous paper 
\cite{vonNeumann:1951}. In this paper we generalise another procedure set 
out in the paper - rejected sampling. 

As we mentioned earlier the initial sample should be prepared with an 
importance sampling function, in order to increase selection efficiency. 
Since we are interested in an automated procedure, we will use the VEGAS 
algorithm, the de facto standard of importance sampling in HEP. It is worth 
noticing that VEGAS itself can be improved, as is described in Ohl's 
paper~\cite{Ohl:1998jn}. For some practical and historical reasons we will 
apply the slightly adapted VEGAS algorithm from CompHEP~\cite{Pukhov:2003}. 

In Sec.~\ref{def} we give some mathematical arguments in favour of the 
iterative von Neumann procedure and formulate the method itself. In 
Sec.~\ref{stopdef} we define a stopping rule for the method. Three model, 
but rather realistic, examples are considered in Sec.~\ref{examples}. 
Final remarks and conclusions are given in the section~\ref{concl}. 

\section{Iterative von Neumann procedure}
\label{def}
We start from the standard ``hit-and-miss'' von Neumann procedure applied 
to a non-negative function $f(\bar{x})$, where $\bar{x} = \{x_1,...,x_d\}$ 
is a point in the $d$-dimension phase space. For simplicity we limit ourselves 
to a bounded domain $\bdm$ in the phase space. At first, we prepare a sample 
of $N$ phase space points $\{\bar{x}_i\}_N$, $i = 0, ..., N-1$ in $\bdm$, 
distributed according to an importance sampling distribution $g(\bar{x})$, 
where $g(\bar{x})$ is a positive function normalised with the condition 
$\int_{\bdm} g (\bar{x}) d\bar{x} = 1$. So, the corresponding weight $\omega_i$ 
of each point is equal to $f(\bar{x}_i)/g(\bar{x}_i)$. Having the weights 
we can estimate the total integral $$I_{tot}=\int_{\bdm} f(\bar{x}) d\bar{x}
\approx\langle f\rangle_{\bar 1} = \langle f/g\rangle_{\bar g} = 
\frac{1}{N}\sum^{N-1}_{i=0}\omega_i,$$ where we denote an estimation of the 
integral with a sample 
of uniformly distributed points $\langle \cdot\rangle_{\bar 1}$. We can also 
obtain an estimation of the accuracy given by $\sigma^2 = \langle \omega^2
\rangle_{\bar g} - \langle \omega\rangle_{\bar g}^2$ -- the standard deviation. 
The next step is to pick out a subset of the point distributed according to 
$f(\bar{x})$ from $\{\bar{x}_i\}_N$. Let us formulate the standard rejection 
sampling procedure\footnote{We will assume the 
point $\bar{x}_0$ has the maximum weight $\omega_0=f(\bar{x}_0)/g(\bar{x}_0)$ 
in the sample.}: 

\begin{itemize}
\item 
Generate a random number $0 < \xi < 1$ 

\item 
Compare $W_i=\omega_i/\omega_0$, if $W_i > \xi$ - accept the point to the final 
sample. 

\item 
Repeat the procedure for all points in the sample. 
\end{itemize}
It is obvious that the final sample of accepted points is distributed 
according to $f(\bar{x})$. But we also obtain the second sample of the 
{\it rejected} points. Usually the sample is much larger than the sample 
of accepted points. So, the main question of this note is whether we can 
extract something useful from analyzing the rejected points. Our answer is 
affirmative. Namely, we will construct a procedure, which ``returns'' the 
rejected point sample to (almost) the initial position before the von Neumann 
procedure was applied. Thus, we will build an iterative Monte-Carlo rejection 
sampling. 

The possibility of the procedure is based on two observations. First, we can 
apply the ``hit-and-miss'' algorithm for points distributed according to 
any importance sampling function. The only conditions we require is our 
ability to calculate the function at each point in the initial sample and 
the positivity of the function, since $g(\bar{x}_i)$ goes to denominator. 
Certainly, the acceptance efficiency will be strongly dependent on the chosen 
function. The second observation (or, strictly speaking, a limitation) is 
the relevance of the maximum value of $f(\bar{x})$ at $\bar{x}_0$. If we 
take $\omega_0$ for the ``hit-and-miss'' procedure the point will be accepted 
by construction, since the acceptance probability of the point is equal to 
1.0. So, the rejected points will describe the region near $\bar{x}_0$  worse 
then any other region, and further samples constructed at the next rejection 
steps will become worse and worse. Thus, we must formulate a simple criterion 
which ensures a sufficiently reasonable description of the region around the 
``maximum'' in the new accepted sample and stops iteration if the description 
becomes ``too poor'' (see more details of the criterion in the next section). 
Therefore, our goal is twofold. First, we must construct a distribution 
function for the rejected points (we make it in this section). After that, 
we have to see whether the rejected points are still able to describe 
$f(\bar{x})$ without serious drawbacks, especially in the region around 
$f_{max}$. 


The first problem can be solved by considering the densities of points in 
three samples: a) the initial one, b) the sample of accepted points after 
the ``hit-and-miss'' procedure, and c) the sample of rejected points. The 
densities are equal to 
$$
\rho_{ini}(\bar{x}) = N \cdot g (\bar{x})\equiv N \cdot g_0 (\bar{x}), 
$$
$$
\rho_{acc}(\bar{x}) = \alpha \cdot N_{acc} \cdot f (\bar{x}), 
$$
$$
\rho_{rej}(\bar{x}) = N_{rej} \cdot g_1 (\bar{x}),
$$
where we assume the importance sampling functions are normalised to 1.0 and 
$\alpha$ is responsible for normalisation of $f (\bar{x})$: $\alpha = 
1 / \int f (\bar{x}) d\bar{x}\approx 1/I_{tot}$. $N_{acc}$ and $N_{rej}$ are 
the numbers of accepted and rejected points correspondingly and $N = N_{acc} + 
N_{rej}$. Our first problem is to 
find the function $g_1(\bar{x})$. Since the samples of accepted and 
rejected points are obtained from the initial sample we can make use of 
the following key equation: 
\begin{equation}\label{dens_balance}
\rho_{ini}(\bar{x}) = \rho_{acc}(\bar{x}) + \rho_{rej}(\bar{x})
\end{equation}
Substituting our expressions for the densities we obtain 
$$
g_1 (\bar{x}_i) = (N  - \alpha \cdot N_{acc} \cdot \omega_i) g_0 
(\bar{x}_i)/ N_{rej},
$$
where $\omega_i$ is the initial weight of the i-th point (for the importance 
sampling function $g_0 (\bar{x})\equiv g (\bar{x})$). 
Since the proper weights of the rejected points distributed according to 
$g_1(\bar{x})$ are $f(\bar{x}_i)/g_1(\bar{x}_i)$ we can obtain the final 
formula for the weight of the point $\bar{x_i}$ (if it has been rejected): 
\begin{equation}\label{key_form}
\omega'_i = \frac{(1 - \epsilon)\cdot\omega_i} 
{1 - \epsilon\cdot\omega_i/ I_{tot}},
\end{equation}
where $\omega_i$ is the point weight {\it before} rejection sampling, 
$\omega'_i$ is its weight according to the distribution $g_1(\bar{x})$, and 
$\epsilon$ is the acceptance efficiency for the sample. 

The first consequence of the formula is that we do not need to calculate the 
new importance sampling function $g_1 (\bar{x})$ for the next iteration at 
all. The new weights of the rejected points are calculated from the old ones by 
means of a very simple function $f(z) = A\cdot z/(1- B\cdot z)$. Thus, we 
come back to the first step and can repeat the von Neumann rejection procedure 
based on the new weights. Obviously, the second iteration will bring less 
points since larger weights became larger and vice versa. 

Accordingly the iterative Monte-Carlo ``hit-and-miss'' procedure has a very 
simple formulation: 

\begin{enumerate}
\item 
Prepare a sample of points $\{\bar{x}_i\}_N$ with weights 
$\omega_i=f(\bar{x}_i)/g(\bar{x}_i)$ and estimate $I_{tot}$

\item 
Perform the von Neumann procedure for the sample. 

\item 
Re-calculate the weights of the rejected points according the formula~(\ref{key_form}) 
and find the point with the maximum weight. It will be the next $\bar{x}_0$.

\item 
Check whether the sample satisfy a stopping condition. If not, repeat the 
iterations starting from point 2.

\end{enumerate}

It is interesting to mention that the transformation~(\ref{key_form}) is 
stable during iterations. If we perform $m$ iterations and accept 
$N_{acc}^{(1)}, ..., N_{acc}^{(m)}$ points in the iterations, then we obtain 
$$
\omega^{(n)}_i = \frac{(1 - \epsilon^{(n)})\cdot\omega_i} 
{1 - \epsilon^{(n)}\cdot\omega_i/ I_{tot}},
$$
where $\epsilon^{(n)}=\sum^m_{i=1}\epsilon^{(i)}\sum^m_{i=1}N^{(i)}_{acc}/N$. 

Let's estimate the a priori number of events, which would be accepted at the 
(n+1)$\rm ^{th}$ iteration, if we know the information from the previous stage. 
In order to accept 
a point, we compare its weight with a uniformly distributed number $\xi$. 
So, we can say that $\omega_i/\omega_0$ is the probability of the point to be 
accepted. Since the transformation~(\ref{key_form}) does not change the 
maximum point, we can write down 
$$
N_{acc}^{(n+1)} = N_{rej}^{(n)}\frac{I_{tot}}{\omega^{(n),'}_{0,rej}} =
I_{tot}\left(\frac{N^{(n)}_{tot}}{\omega^{(n)}_{0,rej}} -  N_{acc}^{(n)}\right) =
N^{(n)}_{tot}(1/\omega^{(n)}_{0,rej} - 1/\omega^{(n)}_{0,tot}),
$$
where $N^{(n)}_{tot}$, $N^{(n)}_{acc}$, and $N^{(n)}_{rej}$ are the numbers 
of points in the sample before $\rm n^{th}$ iteration, and the numbers of 
accepted and rejected points at the iteration. $\omega^{(n)}_{0,tot}$, 
$\omega^{(n)}_{0,rej}$ and $\omega^{(n),'}_{0,rej}$ are the maximum weight 
in the whole sample before the $\rm n^{th}$ iteration, the maximum weight 
of rejected points only and the point weight calculated according to 
(\ref{key_form}). In the formula we use an approximate equality for the a 
priori and a posteriori acceptance efficiency at the $\rm n^{th}$ iteration: 
$\epsilon_{est} = I_{tot}/\omega_0 \approx \epsilon_{real} = N_{acc}/N$. 

\section{The stopping rule for iterations}
\label{stopdef}
The procedure constructed in the previous section lacks only one element -- 
a reasonable stopping rule. In order to find the condition let us look at 
the formula~(\ref{key_form}) in more detail. It fails to produce sensible 
weights if the function $f(\omega) = 1 - \epsilon\cdot\omega/I_{tot}$ is 
less than or equal to zero. It is clear that the worst case happens if we 
substitute the maximum weight $\omega_0$. Let us show that $1 - \epsilon
\cdot\omega_0/ I_{tot} = 0$ ``on the average''. As we pointed out above, the 
probability of a point to be accepted is equal $W_i = \omega_i/\omega_0$. 
Since any accepted point increases the accepted point sample, the average 
number of accepted point is $N_{acc} = \sum_{i=0}^{N}W_i=N\cdot I_{tot}/
\omega_0$ or $1-\epsilon\cdot\omega_0/I_{tot}=0$. 
So, it looks as the standard von Neumann procedure picks out all possible 
points and already the first re-weighting brings us to an inapplicable 
state of the iterative method. But, the solution of the drawback happens 
automatically, merely due to the definition of the von Neumann procedure. 
By construction, $\bar{x}_0$ {\it always} falls within the accepted point 
sample, and for almost all rejected points the formula~(\ref{key_form}) 
gives a finite answer. It also makes clear the source of the extra events. 
We effectively lower the normalised factor $\omega_0\to\omega_{0,rej} < 
\omega_0$, so there is a room for more accepted events. But the reduction is 
rather smart and does not spoil the statistical sense of accepted events. 
Certainly the rejected events give a cruder estimation of the function and 
in realistic cases there exists a neighbourhood of $\bar{x}_0$ where we do 
not have any rejected points at all. So, we lose some part of the total 
integral. As we see in the next section the total integral in iteration 
tends to decrease and, thus, further iterations give an underestimated 
simulation of our function $f(\bar{x})$. 

Not we can formulate a natural criterion of the stopping condition for the 
iterative procedure. At the beginning we have a sample which gives us an 
estimation of the total integral with some error. It is estimated from the 
standard deviation calculated using the sample. We can continue the 
iterations as long as the integral estimations at the first and $m$-th steps 
$I_{tot}^{(0)}$ and $I_{tot}^{(m)}$ satisfy the following inequality 
\begin{equation}\label{stop_cond}
|I_{tot}^{(0)} - I_{tot}^{(m)}| \le \sigma_0 + \sigma_m,
\end{equation}
where $\sigma_{0,m}$ are the standard deviations at the first and $m$-th 
steps. In this case, we still properly estimate the function $f(\bar{x})$ at 
the $m$-th step, in general, and can extract points distributed according 
to the function. Now we are able to formulate the algorithm completely: 

\begin{enumerate}

\item Prepare a sample of points $\{\bar{x}_i\}_N$ with calculated weights 
$\omega_i=f(\bar{x}_i)/g(\bar{x}_i)$ and estimate $I_{tot}$ 

\item Perform the von Neumann procedure for the sample 

\item Re-calculate weights of the rejected points according the formula 
(\ref{key_form}) and find the point with the maximum weight. It will be the 
next $\bar{x}_0$ 

\item Check whether all the calculated weights are positive. If not, stop 
iteration 

\item Estimate the total integral value and check whether the inequality 
(\ref{stop_cond}) holds true. If yes, repeat the iteration starting from 
point 2, otherwise stop. 

\end{enumerate}

The final problem we come across is the question whether we can mix accepted 
events obtained in the all iterations. In fact, it is easy to prove that 
the accepted events have equal weights $I_{tot}=\langle f\rangle_{\bar 1} = 
\langle\omega\rangle_{\bar g(x)}$. The points are distributed according to 
$g(\bar{x}) = f(\bar{x})/\langle f\rangle$, if we substitute the expression 
to the weight formula $\omega_i=f(\bar{x}_i)/g(\bar{x}_i)=\langle f\rangle$. 
In each step in the algorithm we obtained events with different weights, 
but they are distributed within the sample statistical error and can be 
considered as statistically the same. 

\section{Numerical examples}
\label{examples}

In order to check the practicality of the method we consider three different 
numerical examples of application of the method to complicated 
multidimensional functions. 

The first example is a sum of two Gaussians 
\begin{equation}\label{examp1}
f_1(\bar{x})= A_{norm} \cdot \left[
\exp{\left(-\frac{(\bar{x} - \bar{a}_1)^2}{2\cdot\sigma_1^2}\right)} + 
\xi_6\cdot\exp{\left(-\frac{(\bar{x} - \bar{a}_2)^2}{2\cdot\sigma_2^2}\right)}
\right]
\end{equation}
in a 6-dimensional unit hypercube. For calculations we chose the following 
values of the parameters: $A_{norm}=1.0$, $\xi_6=729.0$, $\sigma_1=0.06$, 
$\sigma_2=0.02$, $\bar{a}_1=(0.2,0.2,0.2,0.2,0.2,0.2)$, and 
$\bar{a}_2=(0.7,0.7,0.7,0.7,0.7,0.7)$. We chose $A_{2}=729.0$ in order to 
compensate for the difference in the Gaussian widths. 

At first, we prepared the VEGAS grid (10 iterations of the grid customization 
were done) for the function and generated the initial sample of points; it 
contained approximately $4.5\cdot10^6$ points. After that, the iterative von 
Neumann procedure was carried out with the sample. By definition, the first 
iteration corresponds to the standard von Neumann selection. On the whole 
we made 25 iterations. No zero or negative weights were calculated in the 
iterations. Fig~\ref{exam1_data} displays the total integrals and cumulative 
numbers of points selected at each iteration. As we see our stopping rule~\ref{stop_cond} is 
satisfied at the $\rm 15^{th}$ iteration. Fig.~\ref{exam1_plots} illustrates 
the quality of fitting of the two obtained samples. We fit the histogram 
$d N_{point}/d x_0$ with the original function (certainly, in 1 dimension). 
The upper plot corresponds to the first iteration sample only (the standard procedure), 
and the distribution for the sample of points combined for fifteen iterations 
is depicted in the lower plot. The latter sample is 6.5 times bigger then 
the former sample. Although the quality of fit for the iterative sample is slightly 
lower, all parameters of the original function are reconstructed better than 
for the standard sample. So, we can conclude the built iterative method works 
properly in this particular example. 
\FIGURE{
\centerline{
\epsfxsize=0.95\textwidth\epsfbox{{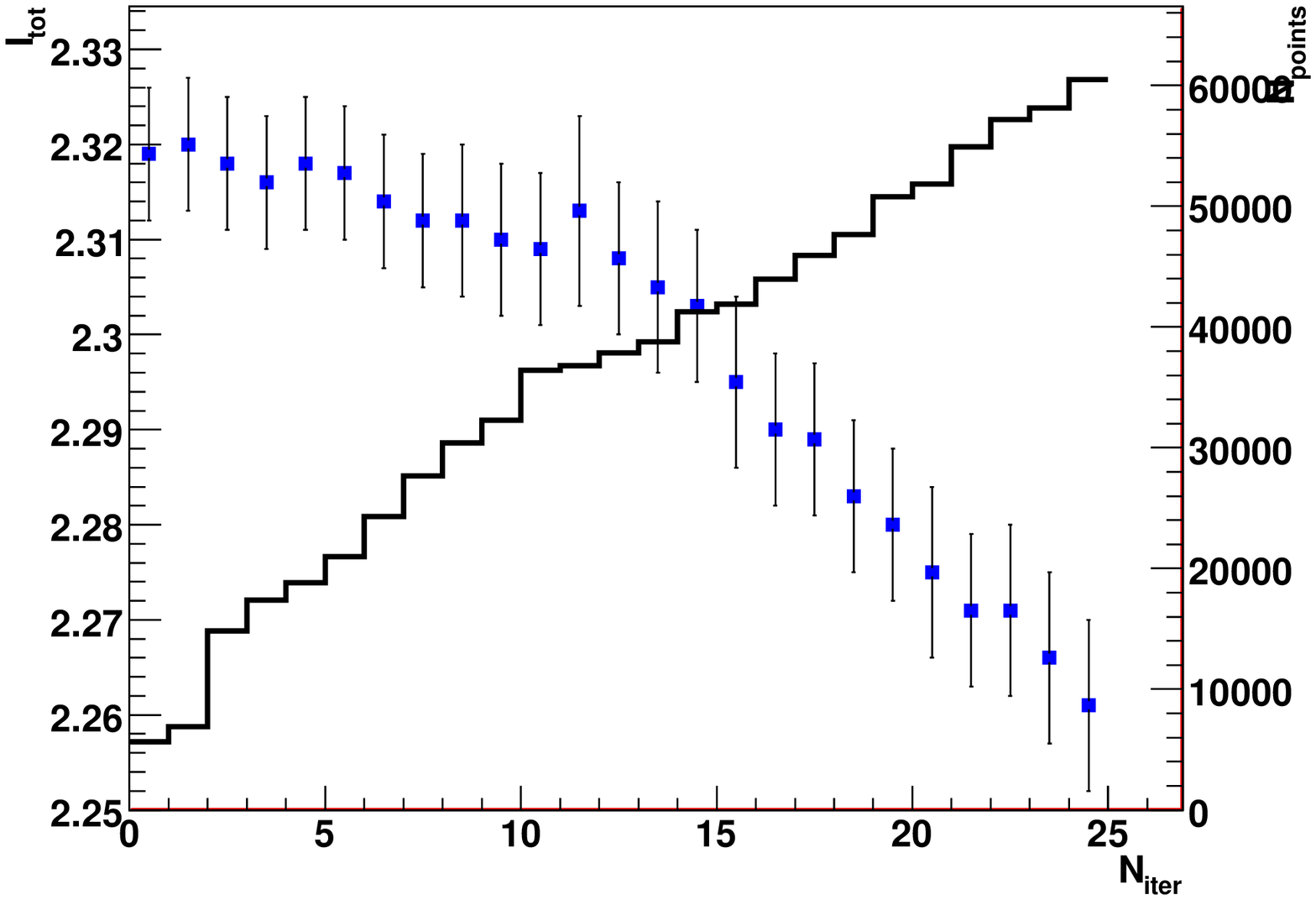}}
}
\vspace{0.3cm}
\caption{ 
Values of the total integral and the cumulative number of selected events at 
iterations of the iterative von Neumann procedure for the function~(\ref{examp1}). 
}
\label{exam1_data}
}

\FIGURE{
\centerline{
\epsfxsize=0.95\textwidth\epsfbox{{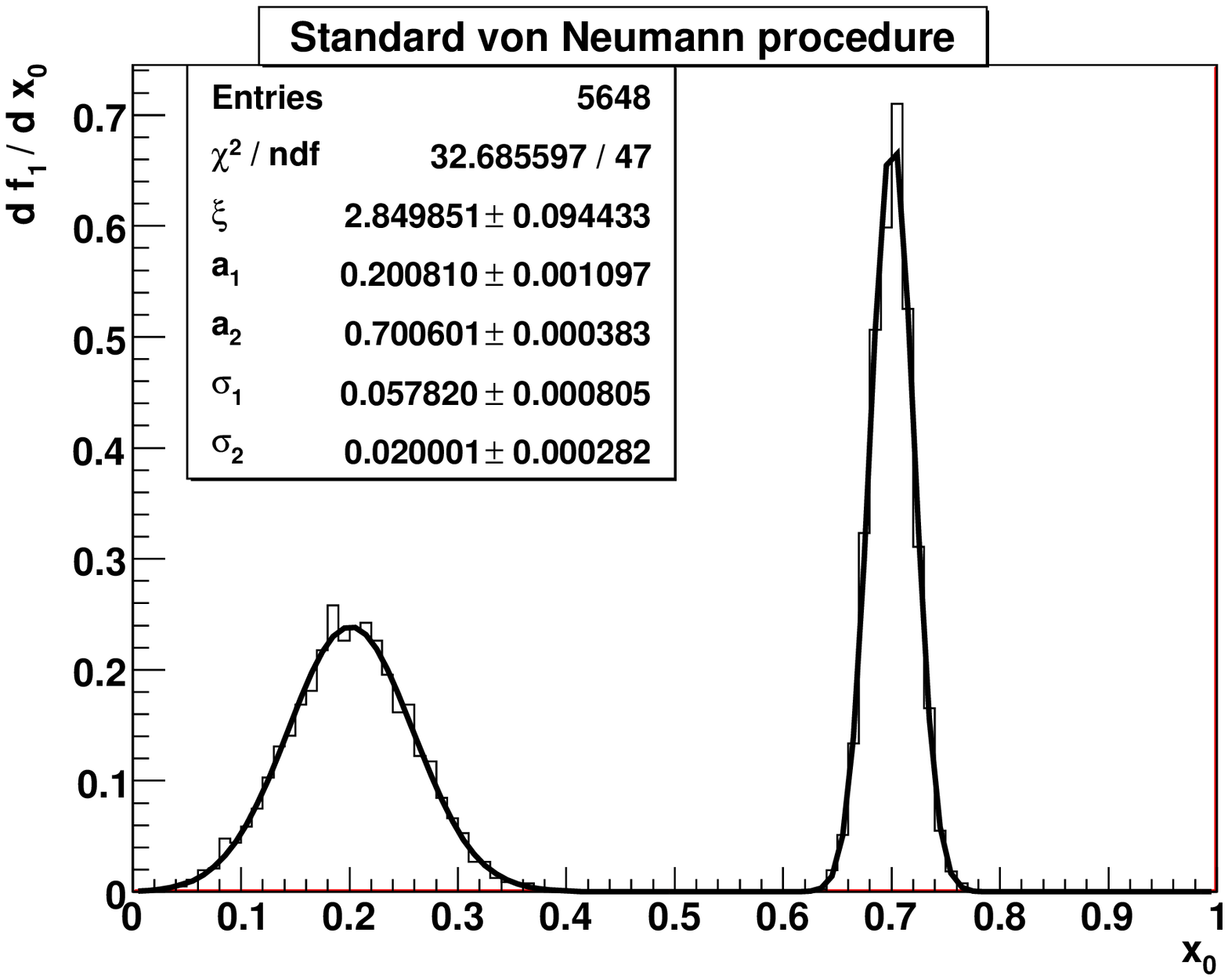}}
}
\centerline{
\hspace{-0.5cm}
\epsfxsize=0.95\textwidth\epsfbox{{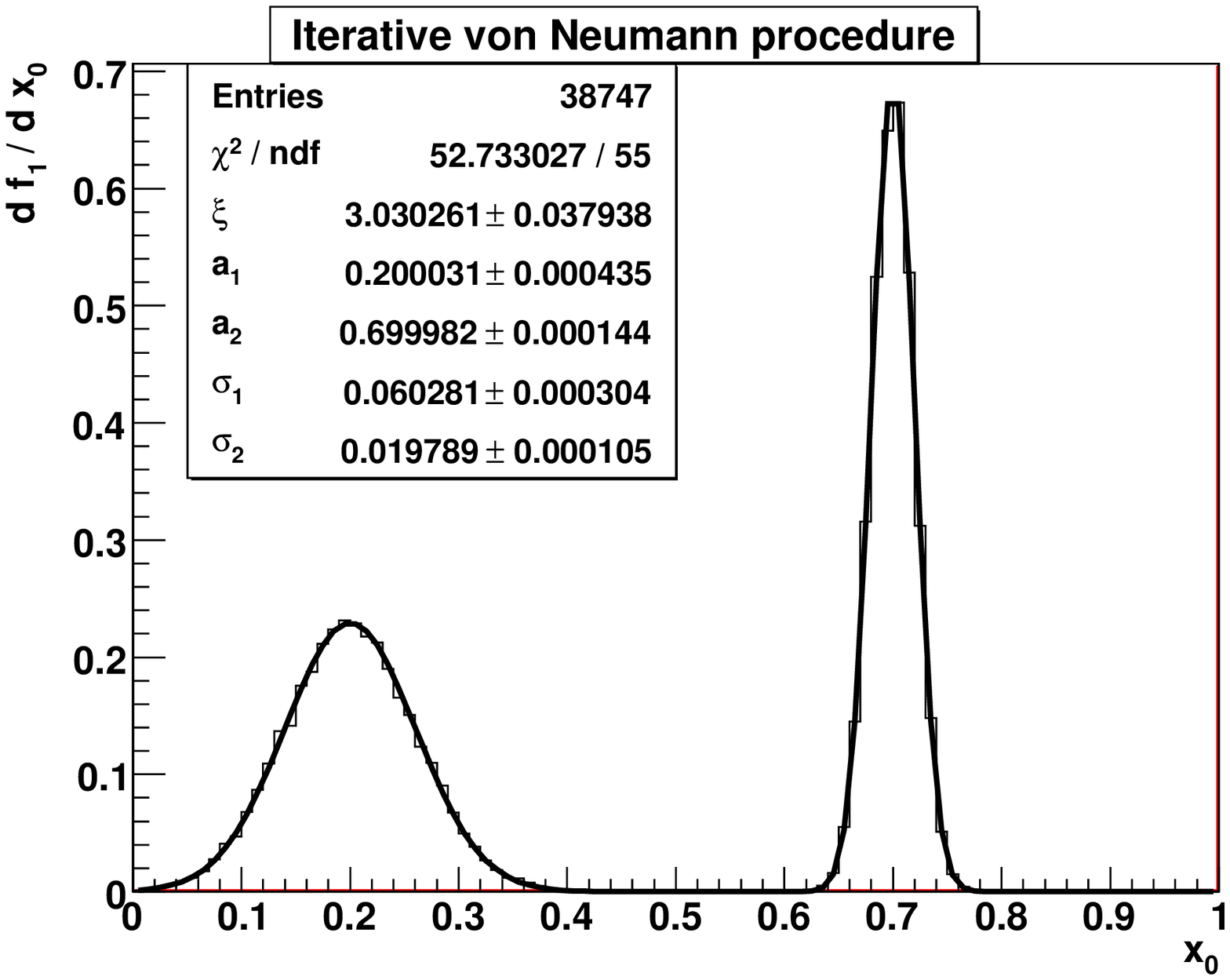}}
}
\caption{ 
The $d N_{point}/d x_0$ distribution for the example~(\ref{examp1}) for two 
approaches, the standard von Neumann procedure (upper plot) and the iterative 
von Neumann procedure (lower plot). Both curves are fitted with the original 
function (for 1 dimension). It is worth noticing that $\xi=\sqrt[6]{\xi_6}$, 
since this is an one-dimensional histogram. 
}
\label{exam1_plots}
}

The second example is a combination of two Breit-Wigner peaks in the 8-dimensional 
unit hypercube. The exact formula is the following: 
\begin{equation}\label{examp2}
f_2(\bar{x})= A_{norm} \cdot \left[
\frac{1}{(a_1 - Y)^2 + \Gamma^2_1} + 
\frac{\xi}{(a_2 - Y)^2 + \Gamma^2_2}\right],
\end{equation}
where we introduce the resonance variable $Y=\sum_{i=0}^{3}{x}_i$. Again, 
we prepared a sample of points in the phase space. The sample size was 
$2.8\cdot10^6$ points for the function with parameters $A_{norm}=60.0$, 
$\xi=0.167$, $\Gamma_1=0.01$, and $\Gamma_2=0.02$. The two peaks are located at 
the points $a_1=0.2$ and $a_2=0.75$. The resonances are taken for the sum 
of four variables $Y=\sum_{i=0}^{3}{x}_i$ in order to make the function less 
sensitive for the VEGAS customization. By definition, the first iteration of 
the iterative von Neumann procedure gave the standard selection with the 
a posteriori efficiency $\epsilon_{eff}=2.57\%$. After that we repeated the 
procedure 16 times for the re-calculated weights (at each step). The total 
integral values and numbers of selected points are reported in Fig.~\ref{exam2_data}. 
Owing to the stopping rule~(\ref{stop_cond}) we should combine samples of the 
first 6 iterations only. This results in the total cumulative efficiency 
$\epsilon_{eff}=4.41\%$. Again, we fit the histogram $d N_{point}/d Y$ with the 
original formula\footnote{Strictly speaking, for the fitting procedure we must 
multiply the expression~(\ref{examp2}) by $Y^3$ since we prepared the sample 
parametrised with $\{x_0,...x_7\}$, but the resonance variable is the 
combination $Y=\sum_{i=0}^{3}{x}_i$. If we transform the initial 
parametrisation into the most natural one, which contains Y, integration over 
the other re-defined variables gives us an extra factor $Y^{N-1}$ in the numerator. 
In our case, N is equal to 4.}. We see the iterative sample results in a better 
fit, i.e a smaller $\chi^2$ and more precise values of the function parameters. 
\FIGURE{
\centerline{
\epsfxsize=0.95\textwidth\epsfbox{{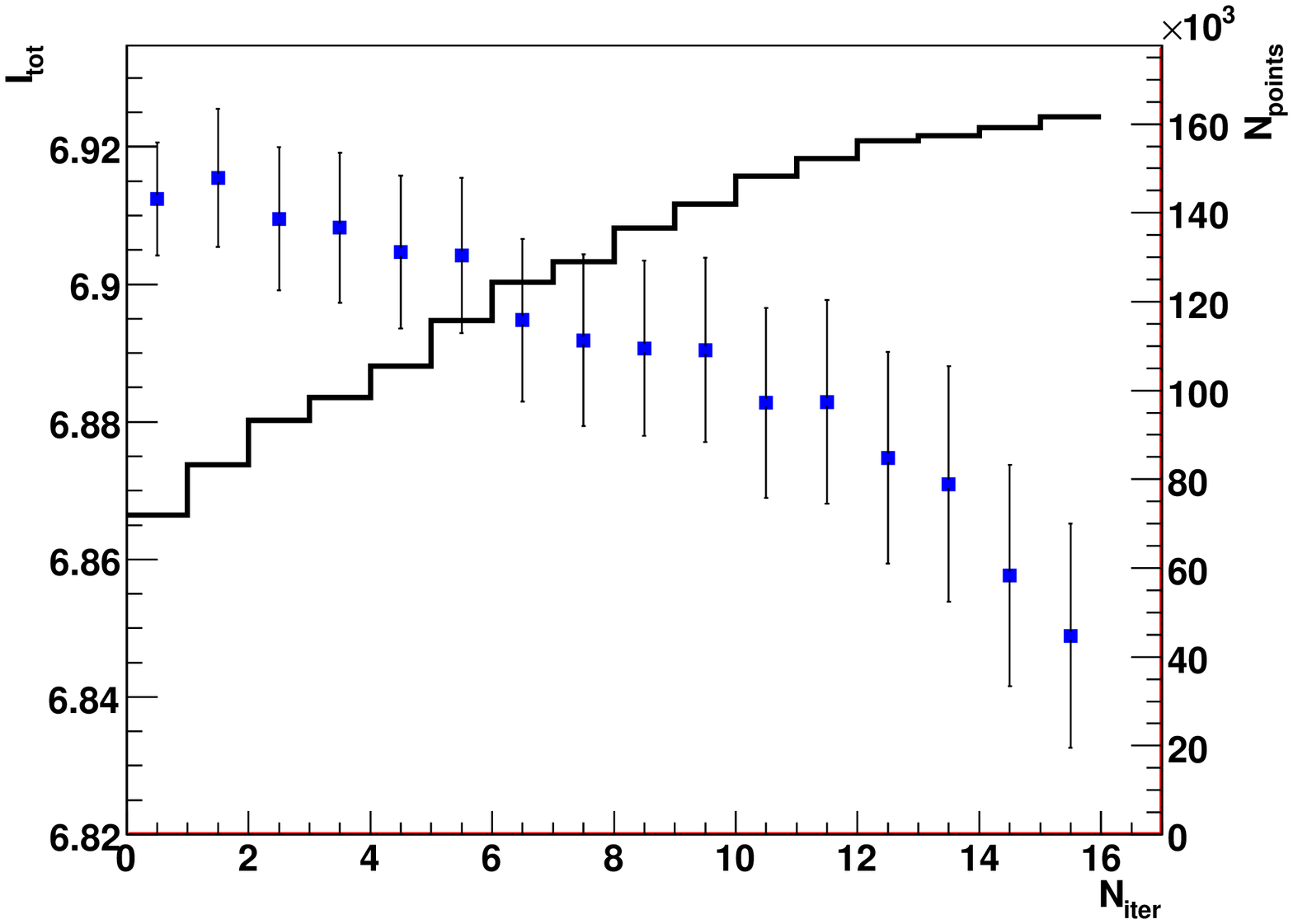}}
}
\vspace{0.3cm}
\caption{ 
Values of the total integral and the cumulative number of selected events at 
iterations of the iterative von Neumann procedure for the function~(\ref{examp2}). 
}
\label{exam2_data}
}
\FIGURE{
\centerline{
\epsfxsize=0.95\textwidth\epsfbox{{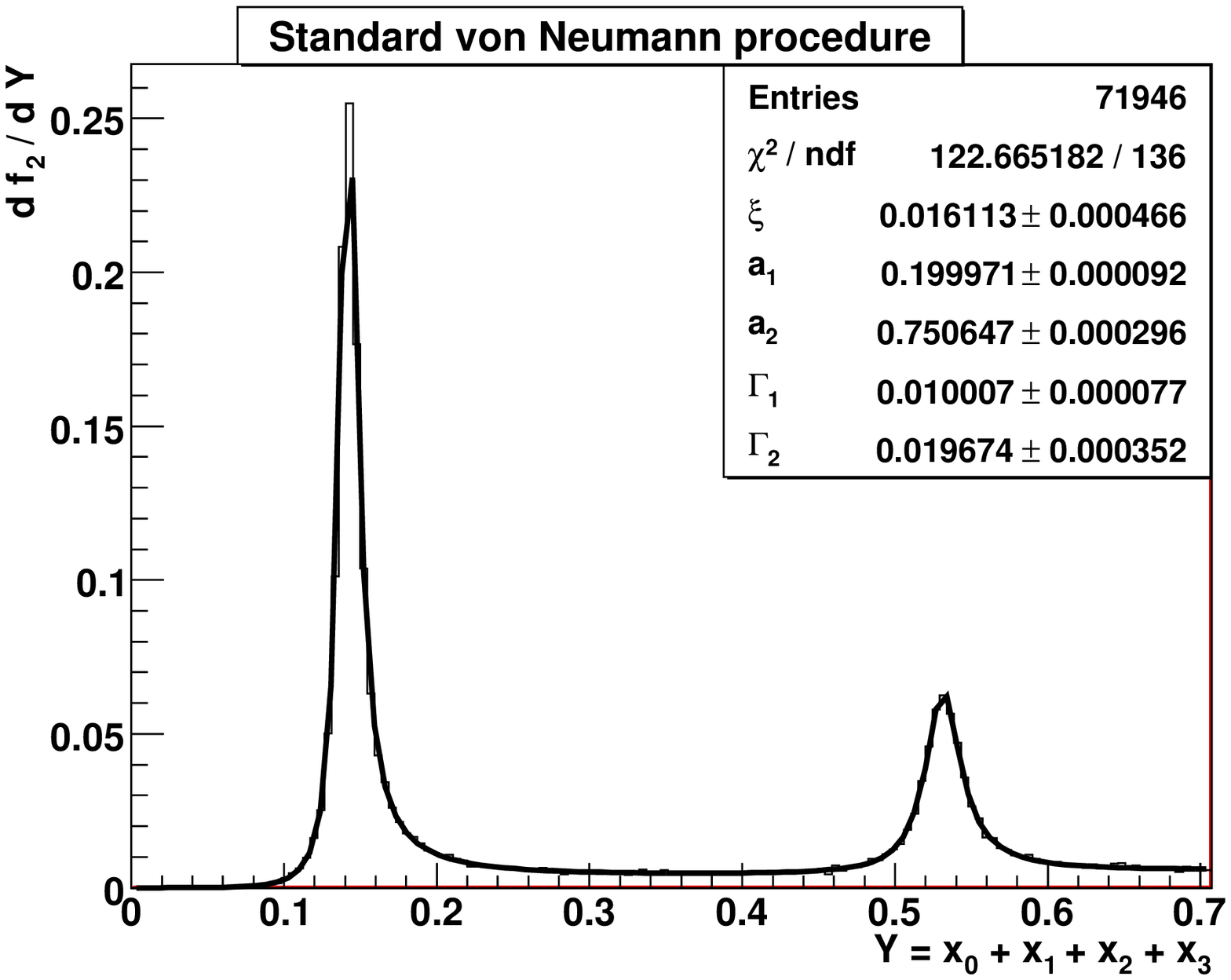}}
}
\centerline{
\hspace{-0.5cm}
\epsfxsize=0.95\textwidth\epsfbox{{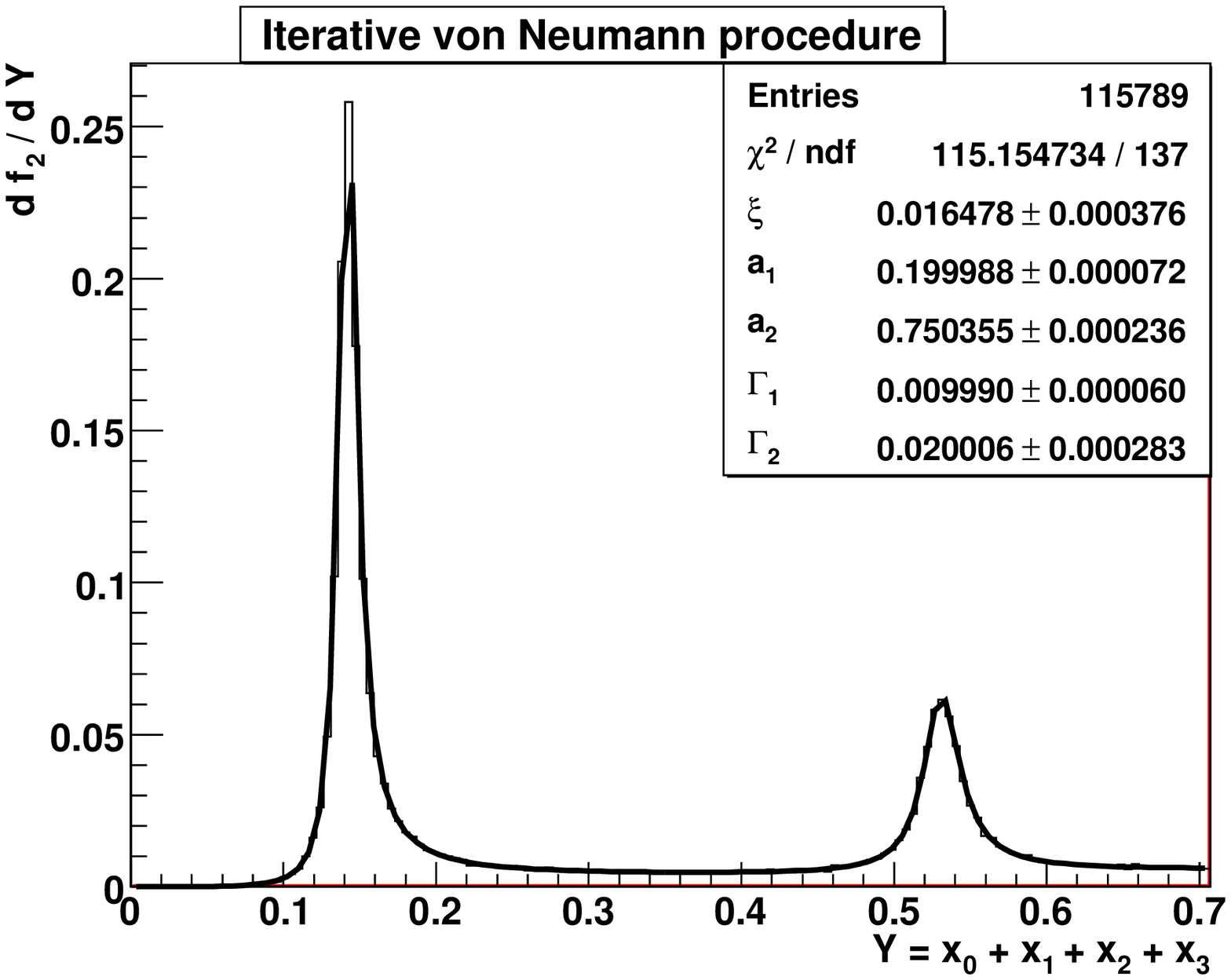}}
}
\caption{ 
The $d N_{point}/d Y$ distribution for the example~(\ref{examp2}) fitted for two 
approaches, the standard von Neumann procedure (upper plot) and the iterative 
von Neumann procedure (lower plot). 
}
\label{exam2_plots}
}

The third example represents a non-symmetric function in a 20-dimensional 
hypercube: 
\begin{equation}\label{examp4}
f_2(\bar{x})= \frac{A_{norm}}{(a_1 + x_0)^{20}}.
\end{equation}
Since $x_0\ge 0$ in the hypercube and the function approaches infinity at 
$x_0=-a_1$, we simulate the right-hand slope of the function, if $a_1>0$. 
We chose the following 
values of the function parameters: $A_{norm}=10^{-55}$ and $a=10^{-3}$. The 
initial sample had $2.23\cdot10^6$ points. The first  (standard) iteration 
gave the a posteriori efficiency $\epsilon_{eff}=3.29\%$. We repeated the 
selection ten times and combined selected points in the final samples for 
the five iterations, according to our sopping rule. The final efficiency is 
equal to $7.28\%$. The total integral values and numbers of selected points 
are reported in Fig.~\ref{exam4_data}. Since we cannot improve the accuracy 
of $A_{norm}$ by our method, we are interested in the quality of the 
reconstruction of $a$ only. We fit the histogram $d N_{point}/d x_0$ with 
the original formula. Again, we see that the iterative sample results in 
smaller $\chi^2$ and more precise value of $a$ (see Fig.~\ref{exam4_plots}). 
\FIGURE{
\centerline{
\epsfxsize=0.95\textwidth\epsfbox{{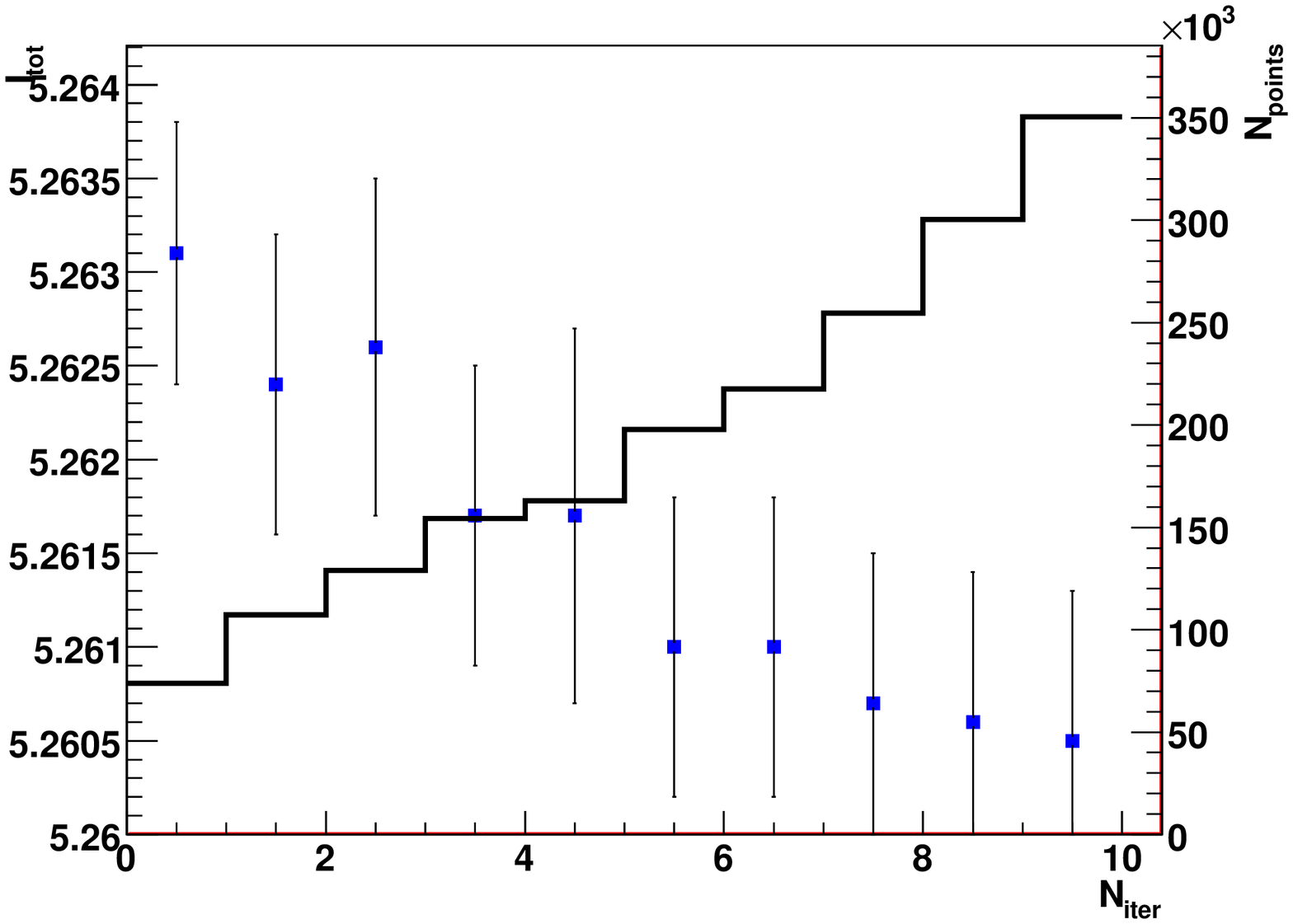}}
}
\vspace{0.3cm}
\caption{ 
Values of the total integral and the cumulative number of selected events at 
iterations of the iterative von Neumann procedure for the function~(\ref{examp4}). 
}
\label{exam4_data}
}
\FIGURE{
\centerline{
\epsfxsize=0.95\textwidth\epsfbox{{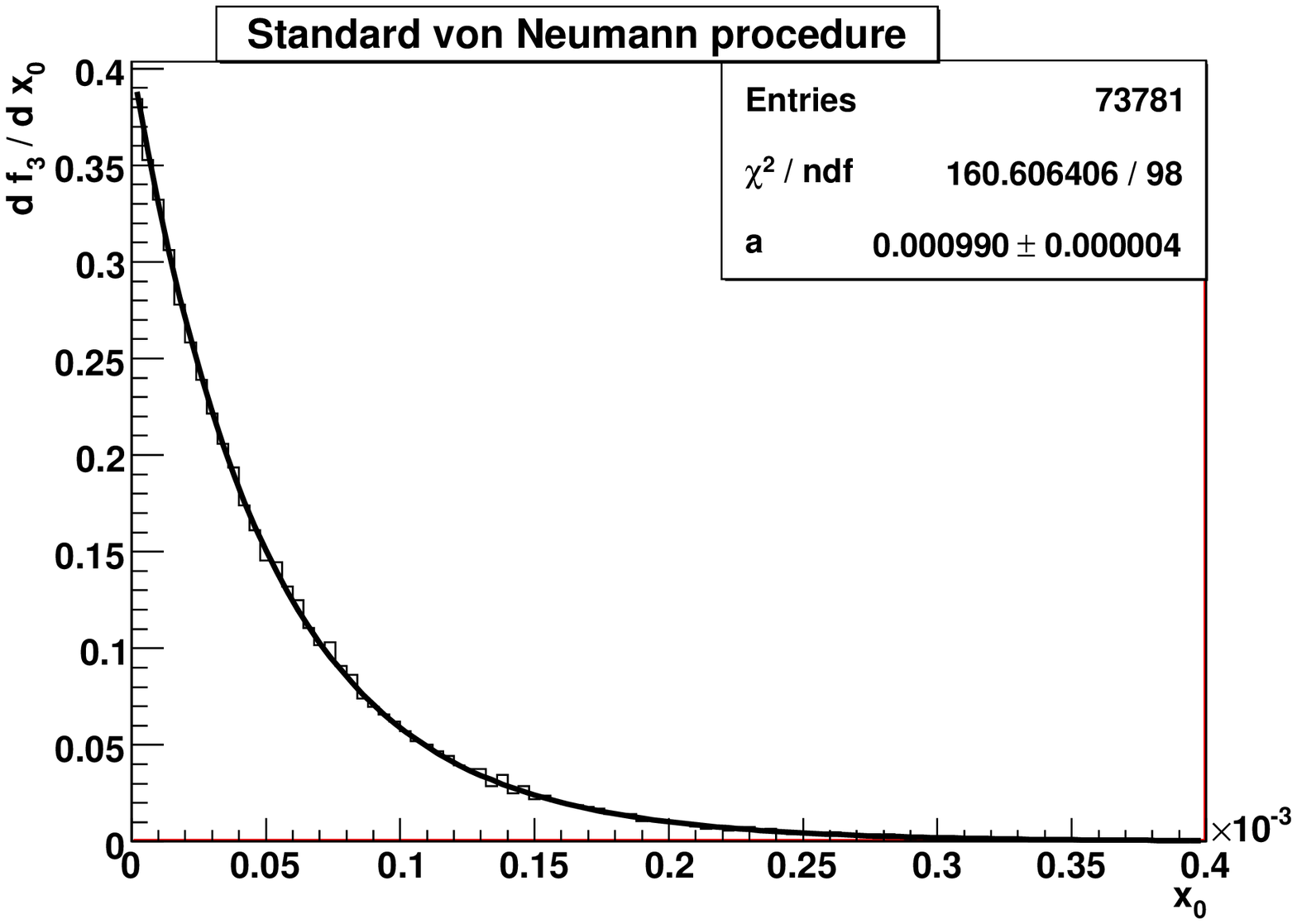}}
}
\centerline{
\hspace{-0.5cm}
\epsfxsize=0.95\textwidth\epsfbox{{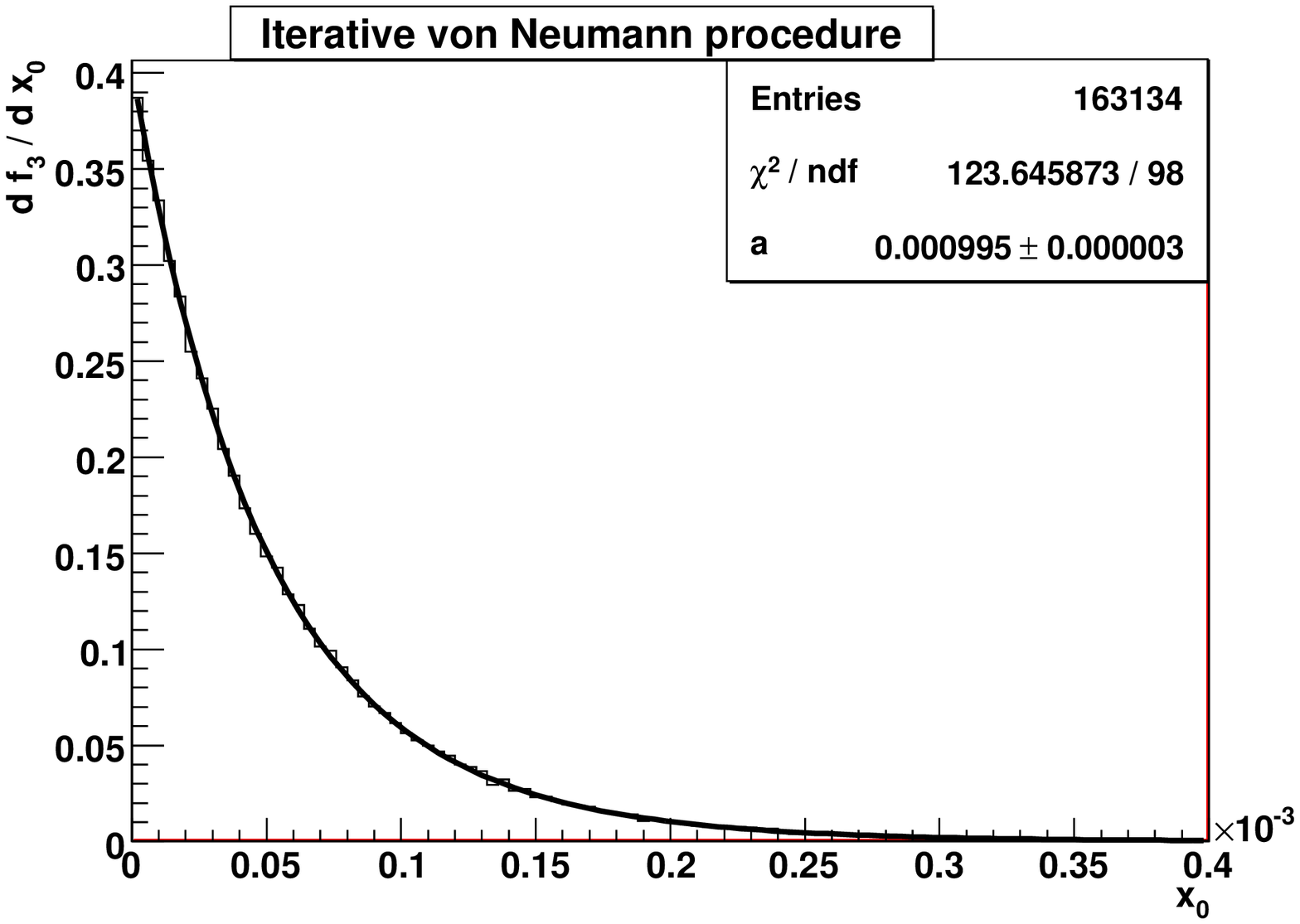}}
}
\caption{ 
The $d N_{point}/d x_0$ distribution for the example~(\ref{examp4}) fitted for two 
approaches, the standard von Neumann procedure (upper plot) and the iterative 
von Neumann procedure (lower plot). 
}
\label{exam4_plots}
}

\section{Discussion and Conclusions}
\label{concl}

A simple generalisation of the standard von Neumann procedure has been 
constructed. The method is based on the observation that in realistic 
Monte-Carlo calculations we always have an uncertainty in the estimation 
of the total integral. This means we can repeat the standard von Neumann 
rejection sampling several times until we are within limits of the uncertainty. 
As soon as we exceed the limit the final sample will start to give worse 
results. Since we exclude some points in the rejection sampling we must 
rearrange weights of points in the rejected sample. We found that the 
transformation is universal (it does not depend on the function considered) 
and extremely simple -- the key formula~(\ref{key_form}). The second obstacle 
which can prevent iteration is negative weights of some points. This can 
happen especially in the case of very low selection efficiency. So, we 
formulated a stopping rule for the iterative procedure, which permits one to 
stay within a well-defined statistical interpretation of selected points and 
to combine all samples obtained in the iterations into one. 

It is worth stressing that, by definition of our method, we cannot improve 
the accuracy of the total integral or, in other words, the quality of the 
reconstruction of the total normalization constant of a function under 
consideration. Moreover, since we select points with larger weights, the 
rejected point sample gives a biased estimation of the total integral. 
Of course we can prevent bad behaviour of the iteration process by 
application of the stopping rule~(\ref{stop_cond}). 

Three numerical examples show the procedure can really bring practical benefits 
in calculations. Owing to the simplicity of the procedure it can have a very broad 
area of application. In fact, almost every sampler, satisfying the condition that 
we should be able to prepare a set of independently generated points with weights, can 
be supplemented with a code producing more points iteratively. The method can be 
easily generalised to stratified sampling, since it can be applied 
independently in each stratum. 

As the first realistic application we are going to implement the method in 
the Monte-Carlo generators CompHEP~\cite{Boos:2004kh} and Herwig++~\cite{Bahr:2008tx}. 
Certainly, the method can be adopted by other popular codes (e.g. PYTHIA, 
MadGraph, Alpgen, Sherpa~)~\cite{other_codes}. 

\section{Acknowledgements}
I would like to thank Bryan Webber and Deirdre Black for useful discussions 
and remarks. I would like to acknowledge the British STFC for the award of 
a Responsive Research Associate position, RFBR (the RFBR grant 07-07-00365-a), 
and European Union Marie Curie Research Training Network MCnet (contract 
MRTN-CT-2006-035606) for partial support of the project. 


\end{document}